\documentclass{PoS}

\title{Recent developments in neutron-proton scattering with Lattice Effective Field Theory}

\ShortTitle{Neutron-proton scattering with NLEFT}

\author{\speaker{Jose Manuel Alarc\'on}\\
        Helmholtz-Institut f\"ur Strahlen- und Kernphysik \& Bethe Center for Theoretical Physics, Universit\"at Bonn, 53115 Bonn, Germany\\
        E-mail: \email{alarcon@hiskp.uni-bonn.de}}

\abstract{In this contribution, we show some recent progress in the study of neutron-proton scattering with Nuclear Lattice Effective Field Theory (NLEFT). 
We present preliminary studies of both, the uncertainties in the $np$ phase shifts extracted with NLEFT, and the lattice spacing dependence in the transfer matrix formalism.
Such investigations have not been performed before in the literature, and will be relevant for Monte Carlo simulations of nuclear structure with NLEFT.}

\FullConference{The 8th International Workshop on Chiral Dynamics, CD2015 ***\\
		29 June 2015 - 03 July 2015\\
		Pisa,Italy}

\begin{document}

\section{Introduction}

The many-body nuclear problem is one of the most important and challenging problems in physics nowadays. 
To understand how nuclear structure emerge from the fundamental principles of the Standard Model would help, not only in bridging the gap between the macroscopic world and the underlying fundamental principles, but also in providing extremely valuable information to many experimental programs. 

One way to tackle this problem, keeping the underlying principles of the Standard Model, is to employ effective field theories (EFT).
In fact, Weinberg showed long time ago \cite{Weinberg:1991um} how to a-pply EFTs to study the NN interaction on chiral symmetry grounds, the relevant symmetry at the range of energies of interest for nuclear physics. 
Chiral EFT has the advantage of i) having the same physical content as the fundamental theory of the strong interactions (QCD), ii) it is systematically improvable, including also many-body forces with the correct hierarchy, and iii) provides a systematic way to assess the theoretical errors. 
This approach, in the one-baryon sector, has proven very successful in studying the properties of the nucleon and fundamental hadronic processes \cite{Bernard:1991rq,Lensky:2009uv,Alarcon:2011zs,Alarcon:2012kn,Alarcon:2012nr,Bernard:2012hb,Alarcon:2013cba,Lensky:2014dda,Blin:2014rpa,Blin:2015nha,Siemens:2014pma}.

Unfortunately, the non-perturvative nature of the NN interaction complicates considerably the EFT approach \cite{Weinberg:1991um}. 
Nevertheless, there are many ways to circumvent this problem.
The one that we apply is to formulate the theory on the lattice.
This idea was exploited long time ago in the pioneering calculation of Ref.~\cite{Borasoy:2006qn}, where it was shown how to study the NN interaction with the NLEFT formalism. Several important calculations of many-nucleon systems followed this one, like {\it ab initio} studies of the Hoyle state \cite{Epelbaum:2011md}, the triple-alpha process \cite{Epelbaum:2012iu} or alpha-alpha scattering \cite{Elhatisari:2015iga}.
In these calculations, the two-body forces are required as an important input, and the systems are simulated using Monte Carlo techniques, in combination with the transfer matrix formalism.

Our aim here is threefold: 1) refine the NN forces improving the extraction of the low-energy constants with more NN information,  2) give, for first time in the NLEFT literature, an estimation of the uncertainties of the approach, 3) study the dependence of the results on the spatial and temporal spacing in the transfer matrix formalism. 
These progress will be very beneficial for future studies of many-nucleon systems with NLEFT.

\section{Formalism}


We put the NN system in a box of size $L^3$, and discretize the spatial direction in steps of $a=1.97$~fm.
Since we use the transfer matrix formalism, we also discretize the temporal direction, in this case, in steps $a_t = 1.32$~fm.
In order to fix the free parameters of the theory, we perform a least squares fit to the phase shifts provided by the Nijmegen group \cite{Stoks:1993tb} including also the binding energy of the deuteron in the evaluation of the $\chi^2$. 
This is defined as follows,

\begin{equation}
 \chi^2 = \sum \Big[  \frac{\delta_\alpha^{latt}(p) - \delta_\alpha^{NPWA}(p) }{\Delta_\alpha(p)}   \Big]  + \Big[  \frac{E_B^{latt} - E_B^{exp} }{\delta E_B^{exp}}   \Big].
\end{equation}

In the previous formula $\delta_\alpha^{latt}$ is the phase shift calculated on the lattice for the channel with quantum numbers $\alpha$, $\delta_\alpha^{NPWA}$ is the phase shift provided by the Nijmegen group, and $\Delta_\alpha$ is the error of the phase shifts, defined as in Ref.~\cite{Epelbaum:2014efa}. 
Regarding the binding energies,  $E_B^{latt}$,  $E_B^{exp}$ and $\delta E_B^{exp}$ are the ones that we calculate on the lattice, the experimental one and its experimental error, respectively.
Notice that this kind of least squares fits has not been done before in NLEFT.

On the other hand, to calculate the phase shift on the lattice, we use the same method as in the Monte Carlo simulations, the so-called "Spherical wall method" \cite{Borasoy:2007vy}.
This method exploits the fact that the phase shift is just a shift in the phase of the interacting solution in the asymptotic region compared to the non-interacting one, to extract them from the discrete energy levels computed on the lattice.
Solving the Schr\"odinger equation in radial coordinates for both, the interacting and non-interacting system, one sees that the radial part has the the following analytic experssions

\begin{itemize}
 \item Non-interacting
\begin{equation}
 r\cdot R(r) = r\cdot j_\ell(p\, r) \stackrel{r\to \infty}{\longrightarrow} \sin(p\,r - \pi L/2)
\end{equation}

 \item Interacting

\begin{equation}\label{Eq:interacting}
 r\cdot R(r) = r\, \cos \delta_\ell(p) \cdot j_\ell(p\, r) -  r\, \sin \delta_\ell(p) \cdot y_\ell(p\, r) \stackrel{r\to \infty}{\longrightarrow} \sin(p\,r - \pi L/2 + \delta_\ell (p))
\end{equation}
\end{itemize}

that show how the phase shift emerges. In the previous equations, $j_\ell$ and $y_\ell$ are the spherical Bessel functions and $p$ the three-momentum of the nucleons in the center-of-mass system. Imposing a rigid wall in the asymptotic region (see the right panel of Fig.\ref{Fig:PhaseShift&SphericalWall}), we force the wave function to vanish, which means that the phase shift must satisfy the following condition at the wall:

\begin{figure}[h!]
\begin{center}
\includegraphics[width=.55\textwidth,angle=0]{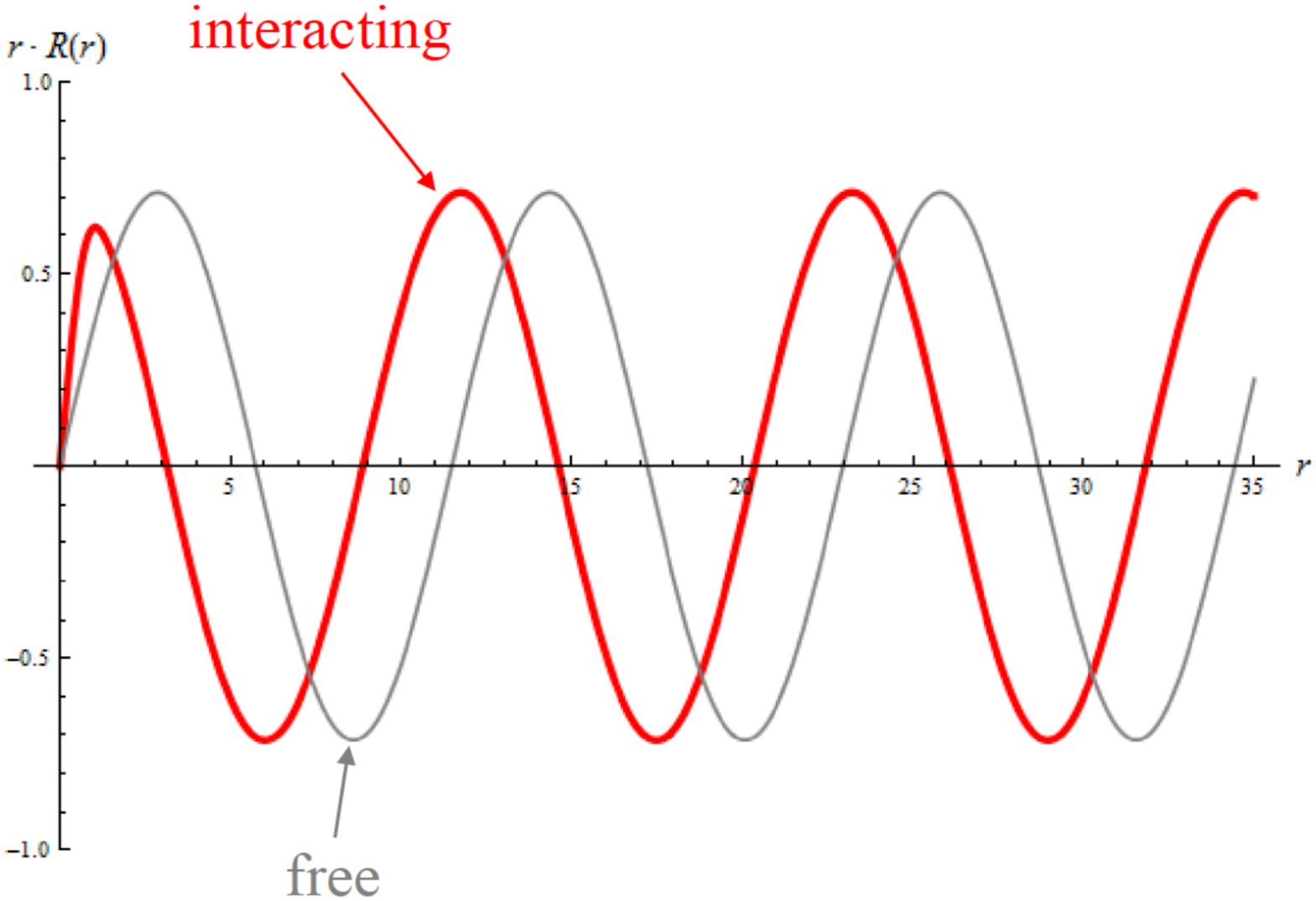}\hspace{1cm}\includegraphics[width=.35\textwidth,angle=0]{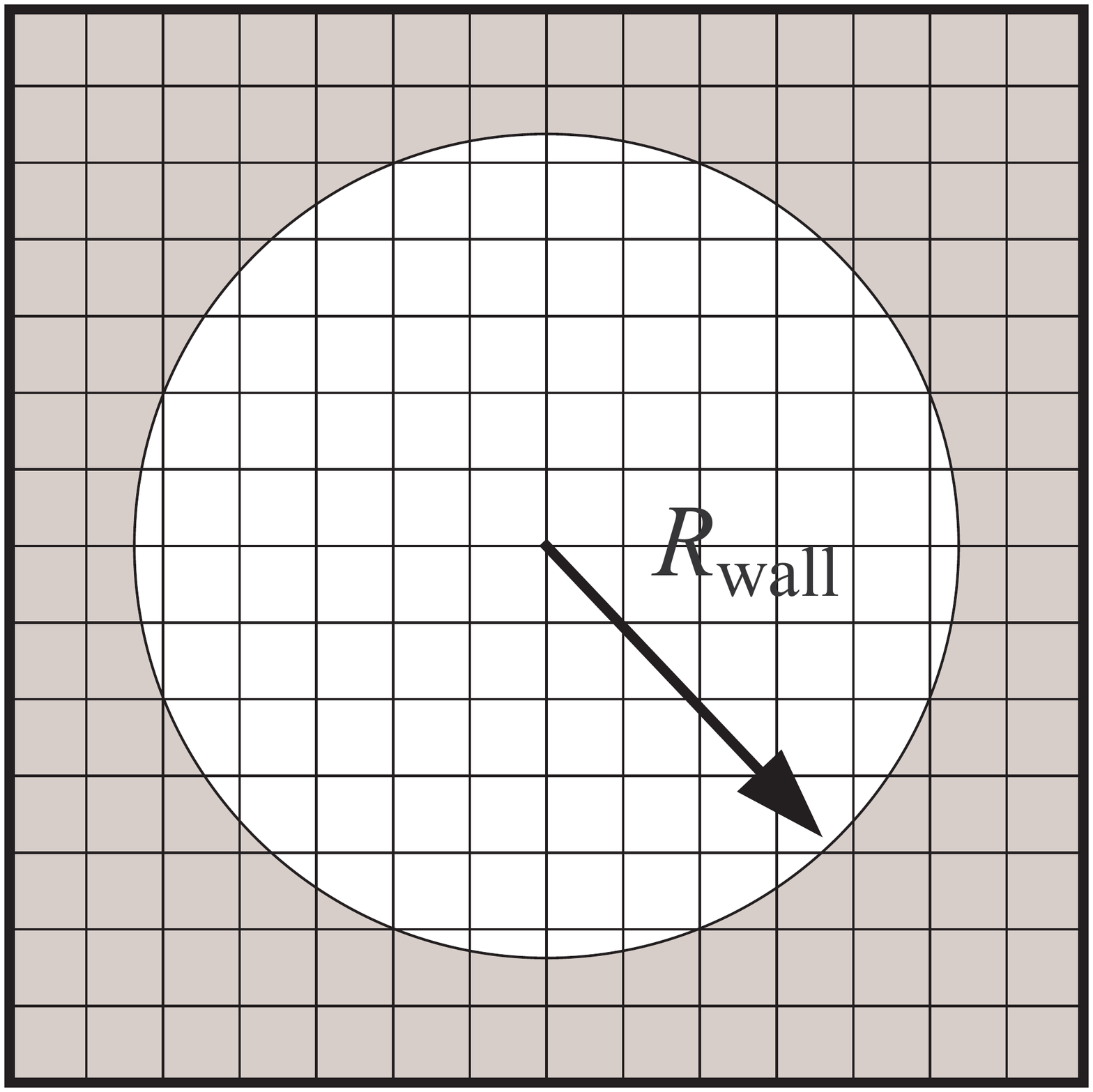}
\caption{Left panel: Visualization of a phase shift in the radial part of the wave function, $R(r)$.
Right panel: Visualization of the spherical wall method. The rigid wall is imposed far away from the interacting region. Outside this wall, grey area, the wave function vanishes. \label{Fig:PhaseShift&SphericalWall}}
\end{center}
\end{figure}

\begin{equation}
\delta_\ell(p)  =  \arctan\Big[ \frac{j_\ell(p\, R_{wall})}{y_\ell(p\, R_{wall})} \Big].
\end{equation}

This condition follows directly from Eq.~(\ref{Eq:interacting}), and allows us to calculate the phase shifts from the energy levels computed from the transfer matrix, since $E=\frac{p^2}{2\mu}$ (being $\mu$ the reduced mass of the NN system).

\section{Results}

\begin{figure}[h!]
\begin{center}
\includegraphics[width=.5\textwidth]{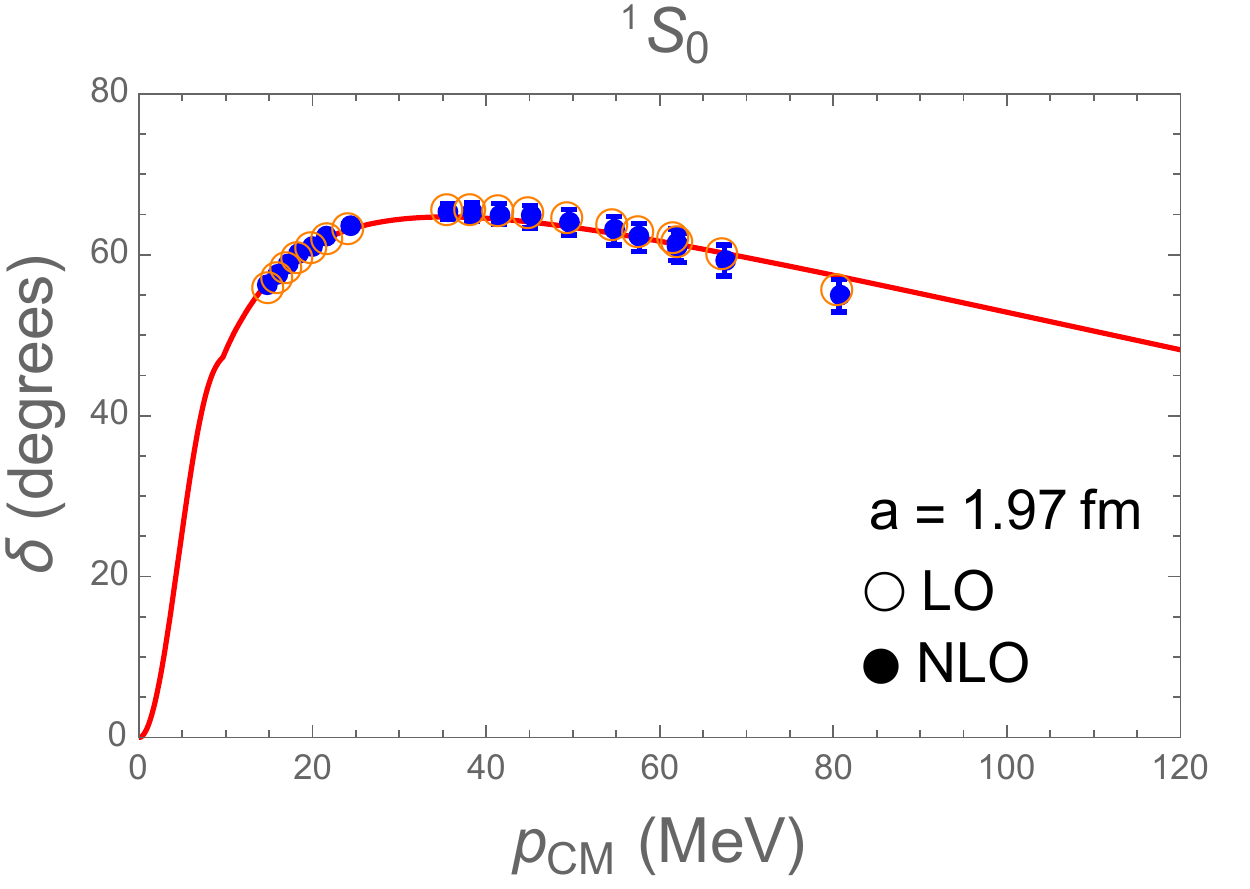}\includegraphics[width=.5\textwidth]{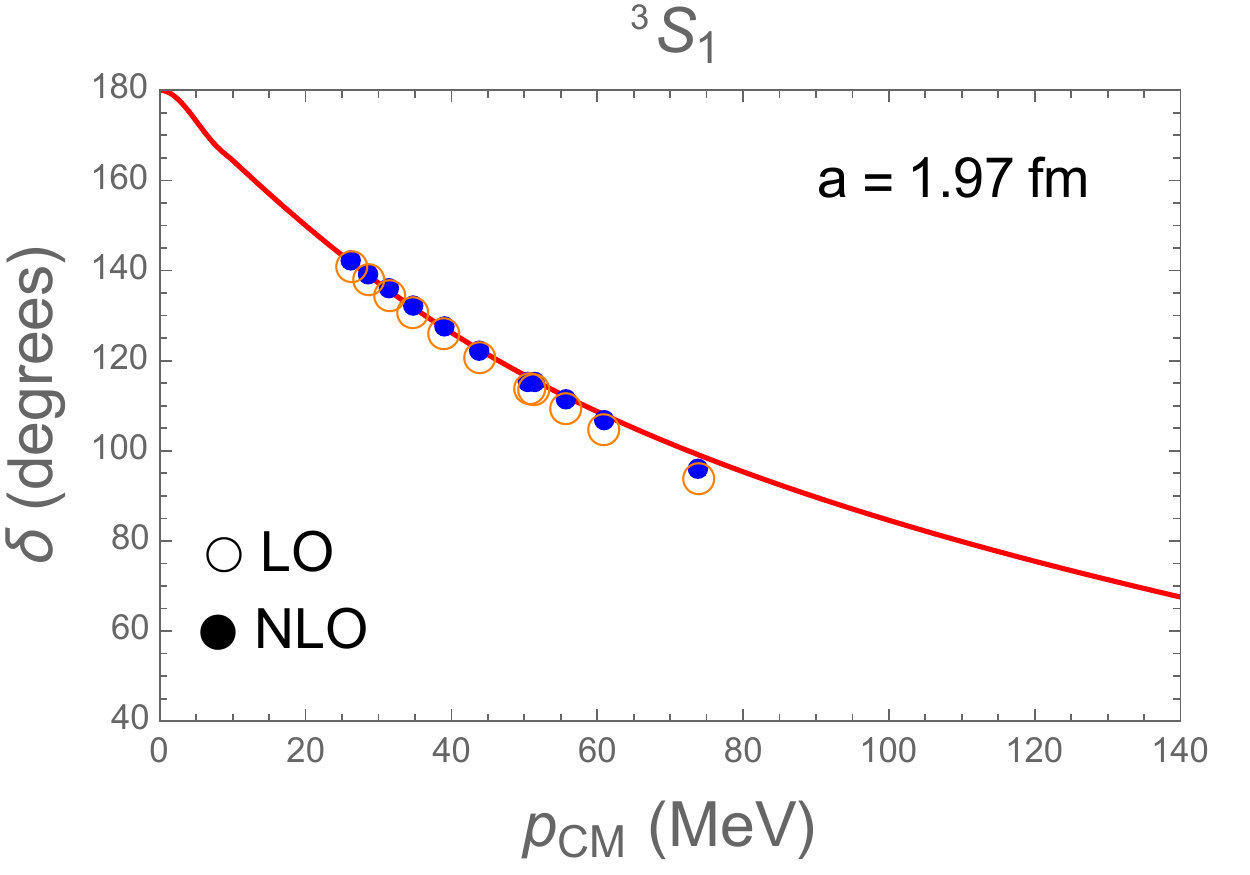}
\caption{Fits to $S$-waves. The red line is the NPWA results, while orange empty circles and blue filled ones are the LO and NLO results, respectively. \label{Fig:Plot_S-waves}}
\end{center}
\label{fig1}
\end{figure}

\begin{figure}[h]
\begin{center}
\includegraphics[width=.5\textwidth]{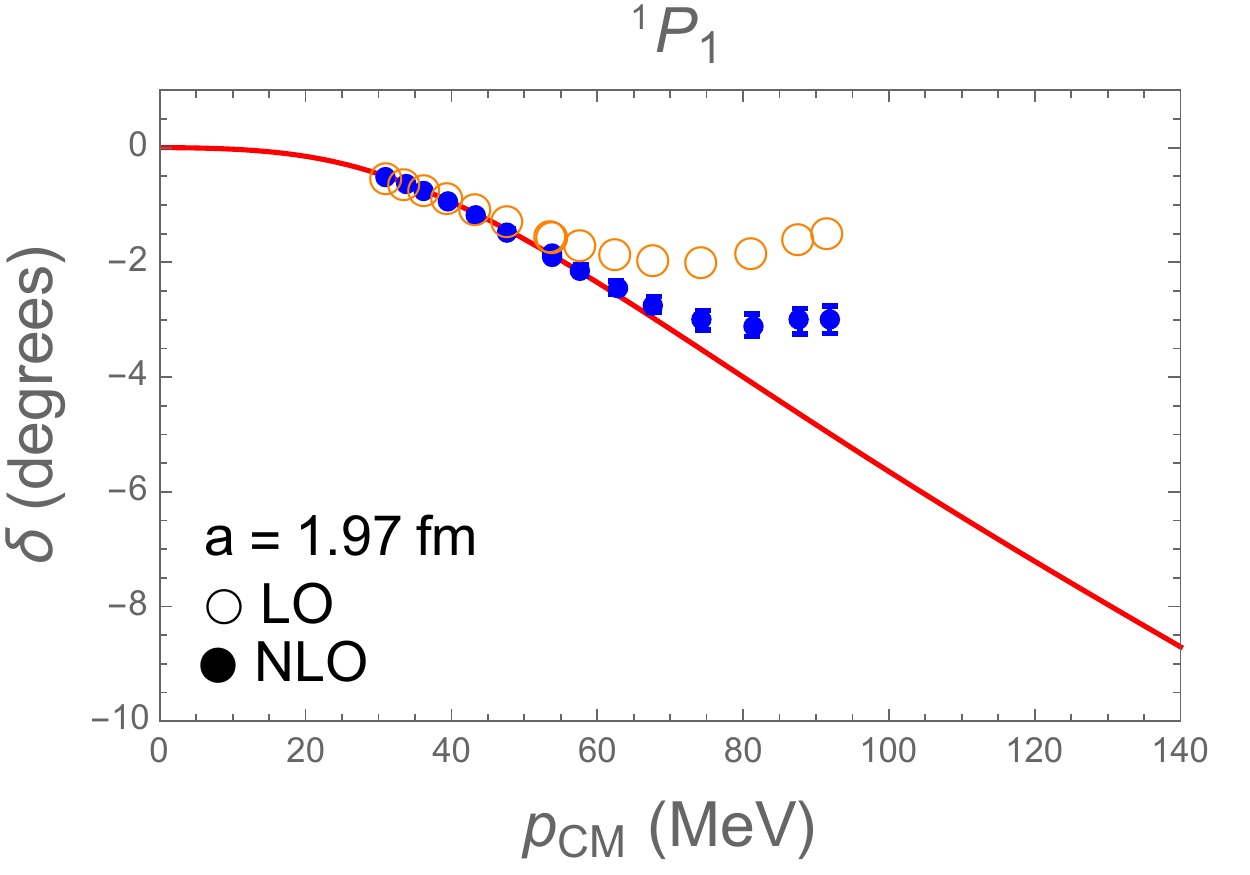}\includegraphics[width=.5\textwidth]{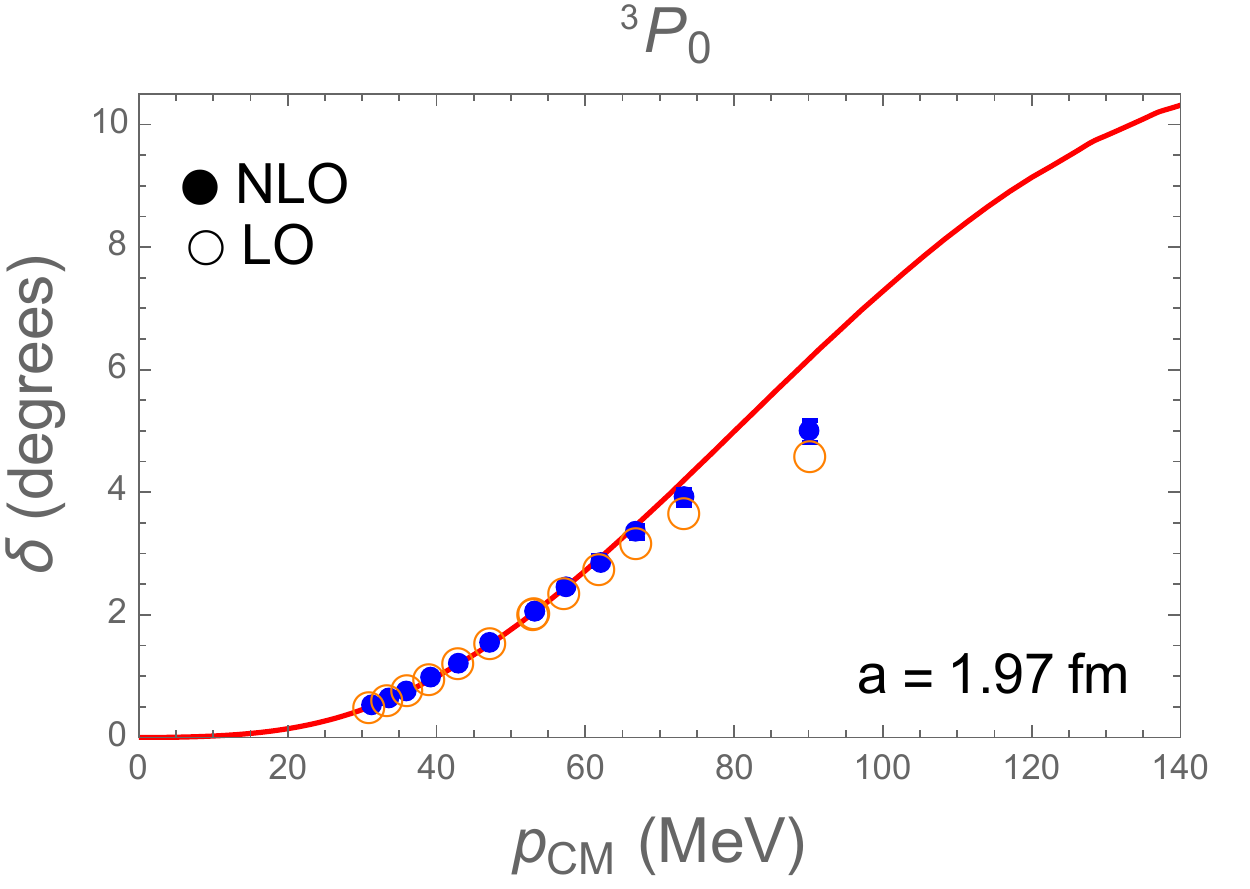}
\includegraphics[width=.5\textwidth]{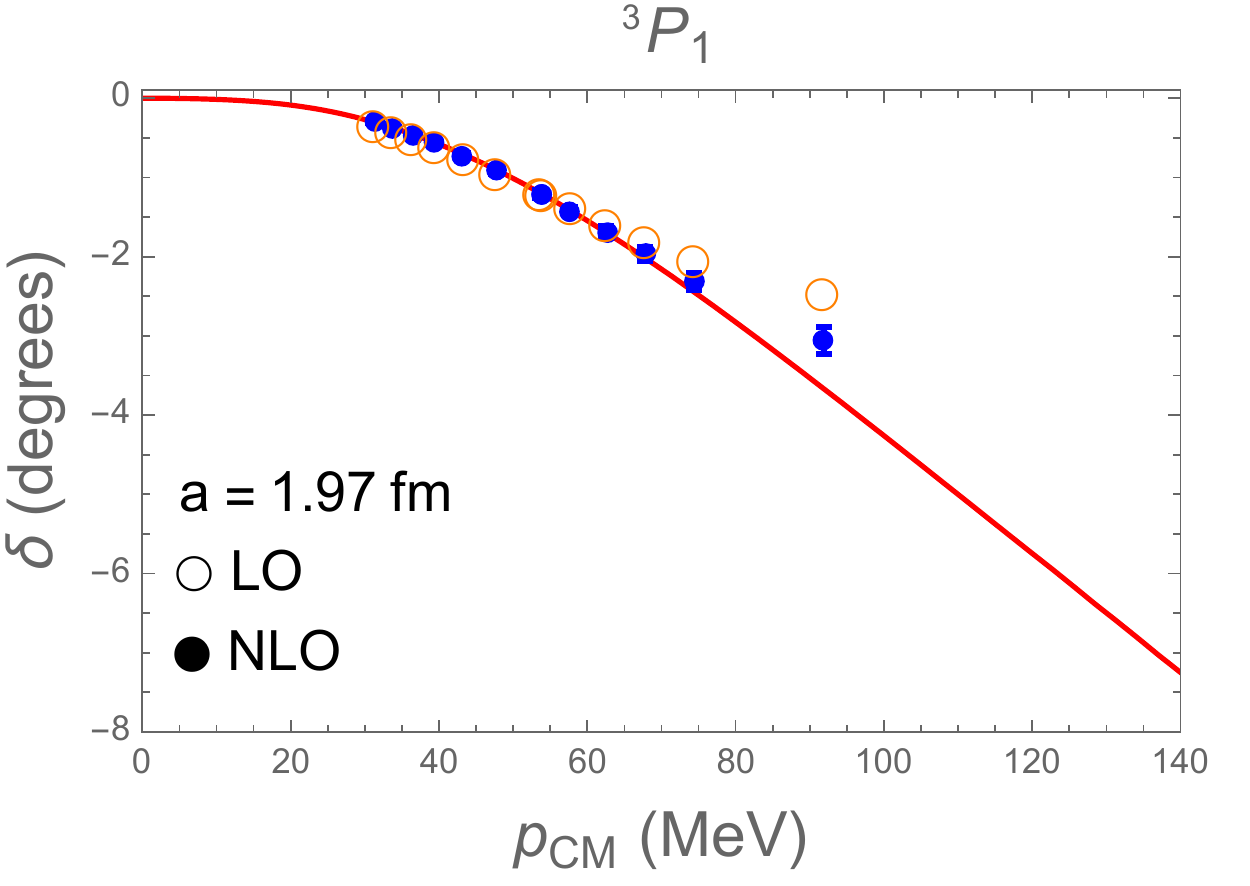}\includegraphics[width=.5\textwidth]{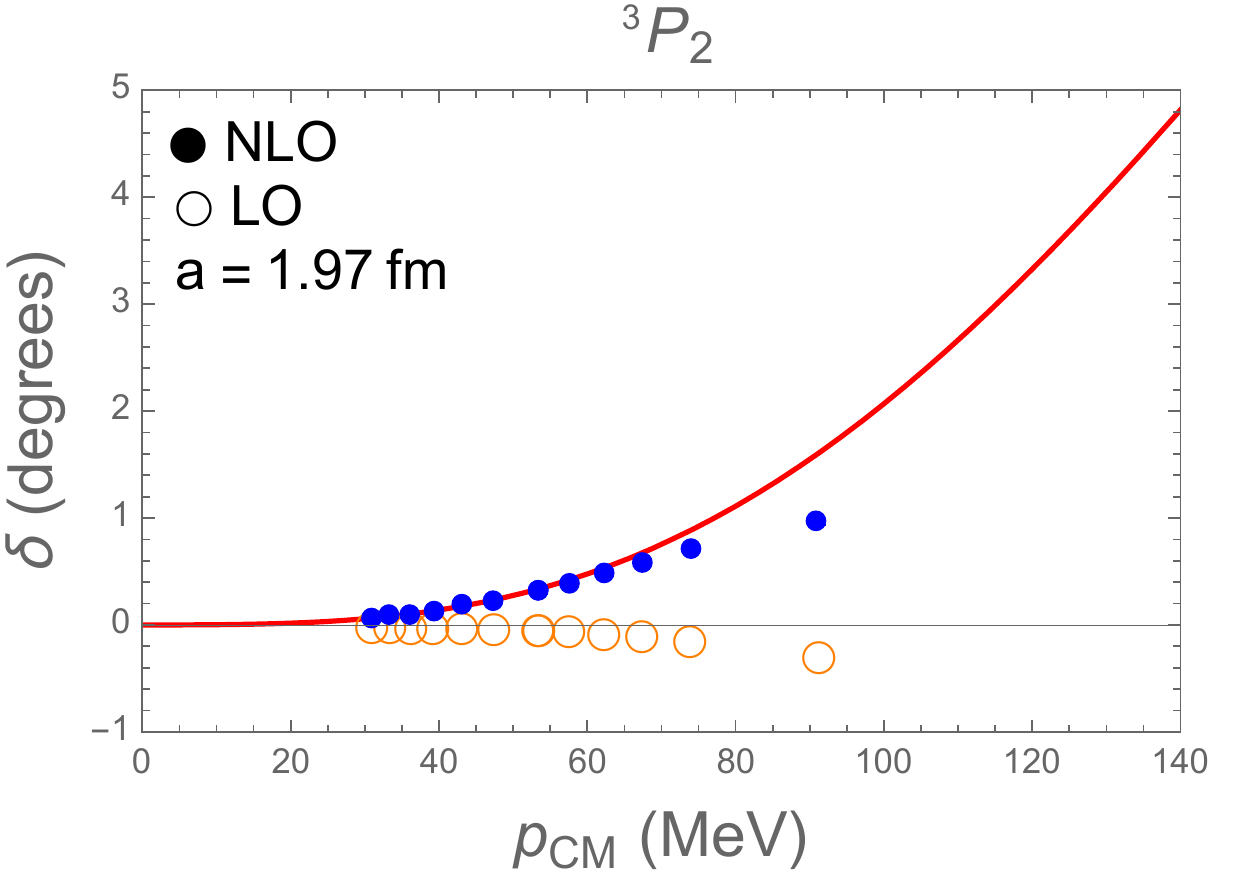}
\caption{Fits to $P$-waves. As before, the red line is the NPWA results, while orange empty circles and blue filled ones are the LO and NLO results, respectively. \label{Fig:Plot_P-waves}}
\end{center}
\label{fig2}
\end{figure}

In order to study the systematics in our analysis, we considered different ranges of energies in our fits. 
We start fitting the LO  using the phase shifts below $p_{CM}^{max} = 30$~MeV up to $p_{CM}^{max} = 60$~MeV, considering only the $S$-waves. 
Then, we continued adding the NLO action and considering also the $P$-waves in a range of energies from $p_{CM}^{max} = 70$~MeV to $p_{CM}^{max} = 100$~MeV for all the channels. 
We perform this two-step-fit because computational limitations made the error analysis of a simultaneous fit of the leading and next-to-leading order unfeasible, due to the large number of free parameters.  
The different ranges considered for the LO and NLO fits give us a range of values for the LECs that translates into uncertainties in the extracted phase shifts.
These uncertainties are reflected in the error bars of Figs.~\ref{Fig:Plot_S-waves} and \ref{Fig:Plot_P-waves}\footnote{Notice that, in some cases, these errors are of the size of the point or even smaller.}.
There is also a second source of uncertainty: the errors from the NPWA phase shfits. 
However, these are completely negligible compared to the systematic errors, which dominate the uncertainties of the fits.
In the following we will show the results of our fits to the $S$- and $P$-waves.

In Fig.~\ref{Fig:Plot_S-waves} we show our results for the $S$-waves.
We see that the LO already gives a very good description at low energies, which translates also in a good description of the threshold parameters (Table~\ref{Fig:Threshold_parameters_S-waves}). 
The NLO correction improves slightly the situation, but it is not so important as in the $P$-waves. 
Regarding the binding energy, already the LO is giving a value close to the experimental one, and the inclusion of the NLO allows to reproduce almost exactly its experimental value.

{\tiny
\begin{table}
\centering
\begin{tabular}{|c|c|c|c|}
\hline
   & LO & NLO & Exp. \\
  \hline
$a_{1S0}$ (fm)  & $-23.31$  & $-23.79$ &$-23.76$ \\
 $r_{1S0}$ (fm) & $2.38$ &$2.57$ & $2.75$\\ 
 $a_{3S1}$ (fm) & $5.26$ & $5.23$  & $5.42$ \\
 $r_{3S1}$ (fm)& $2.05$ & $2.04$ & $1.76$\\ 
\hline
$E_B$~(MeV)   &    -2.223544   &    -2.224574       &  -2.224575(9)        \\
\hline
\end{tabular}
\caption{Threshold parameters for the $S$-waves and the binding energy of the deuteron. \label{Fig:Threshold_parameters_S-waves}}
\end{table}
}

Taking a look at the $P$-waves, Fig.~\ref{Fig:Plot_P-waves}, one realizes that the NLO contribution plays a more relevant role than in the $S$-waves, specially for the $^1 P_1$ and $^3 P_2$.
It is also worthy to point out that, as in the $S$-waves, one sees that the LO and NLO contributions exhibit the convergent pattern expected in our EFT approach: higher orders play a more relevant role at higher energies. 
This is also reflected in the value of the threshold parameters (Table~\ref{Table:Threshold_parameters_P-waves}), in which the NLO correction becomes more important than in the $S$-waves.

{\tiny
\begin{table}
\centering
\begin{tabular}{|c|c|c|c|}
\cline{1-4} 
  \multicolumn{4}{|c|}{Threshold parameters} \\
\cline{1-4} 

   & LO & NLO & Phen.{*} \\
  \hline
$a_{1P1}$ (fm$^3$) & $3.79$ & $2.89$ & $2.81$ \\ 
$r_{1P1}$ (fm$^{-1}$)& $-12.95$ & $-6.28$  & $-7.20$ \\ 
$a_{3P0}$ (fm$^3$)& $-3.14$ & $-2.78$ & $-2.56$\\ 
$r_{3P0}$ (fm$^{-1}$)& $6.56$ & $5.36$ & $4.43$ \\ 
 $a_{3P1}$ (fm$^3$)& $1.99$ & $1.63$ & $1.54$ \\ 
 $r_{3P1}$ (fm$^{-1}$)& $-13.57$ & $-9.71$ & $-8.54$ \\
 $a_{3P2}$ (fm$^3$)& $-0.003$ & $-0.33$& $-0.29$ \\
 $r_{3P2}$ (fm$^{-1}$)& $-1823$ & $16.78$ & $-3.34$\\
  \hline
\end{tabular}
\caption{\small Results for the $P$-wave threshold parameters. *Obtained from an effective range expansion fit to the NPWA phase shifts.\label{Table:Threshold_parameters_P-waves}}
\end{table}
}

After this first error analysis, we are currently improving the accuracy (higher orders) our NN calculations, as well as the action that we have been using so far. 
This will allow us to explore higher energies and include higher partial waves for the determination of the LECs.
However, in order to extend the range of energies it is also necessary to increase the cutoff in energies imposed by the lattice discretization, and have a good control over the posible systematic error introduced in the discretization procedure.
In the following section, we will present some results in this direction.

\section{Study of the lattice spacing dependence}

The spatial discretization imposes a cutoff in the maximum three-momentum that one can reach on the lattice.
This is, in our formulation, $p_{max} = \pi/a$.
Therefore, if one wants increase the value of the cutoff in order to allow higher energy modes, it is necessary to reduce the lattice spacing to lower values.
A study of the lattice spacing dependence in the Hamiltonian formalism, was performed in Ref.~\cite{Klein:2015vna}. 
However, in this case we are interested in doing such kind of analysis for the transfer matrix formalism, which involves also the temporal discretization ($a_t$).
Therefore, when reducing the spatial spacing ($a$), we have to consider also $a_t$.
In our study we decided to fix the value of $a_t$ in three different ways: 1) we keep $a_t$ constant (results in blue color), 2) we keep the ratio $a_t/a$ constant\footnote{Notice that this ratio factorizes in the exponential of the transfer matrix when we write the Hamiltonian in lattice units.}  (results in red color), 3) we keep the ratio $a_t/a^2$ constant (results in green color). 
The last condition ensures to have the same small value in the argument of the exponential in the transfer matrix when changing the value of $a$, since $E_{max} \Delta t \propto (\pi /a)^2 a_t \propto a_t/a^2$.
In this study we only include the LO action, and consider the $np$ $S$-waves and binding energy of the deuteron to determine the values of the LECs for the different spacings.

\begin{figure}[h!]
\begin{center}
\includegraphics[width=.5\textwidth]{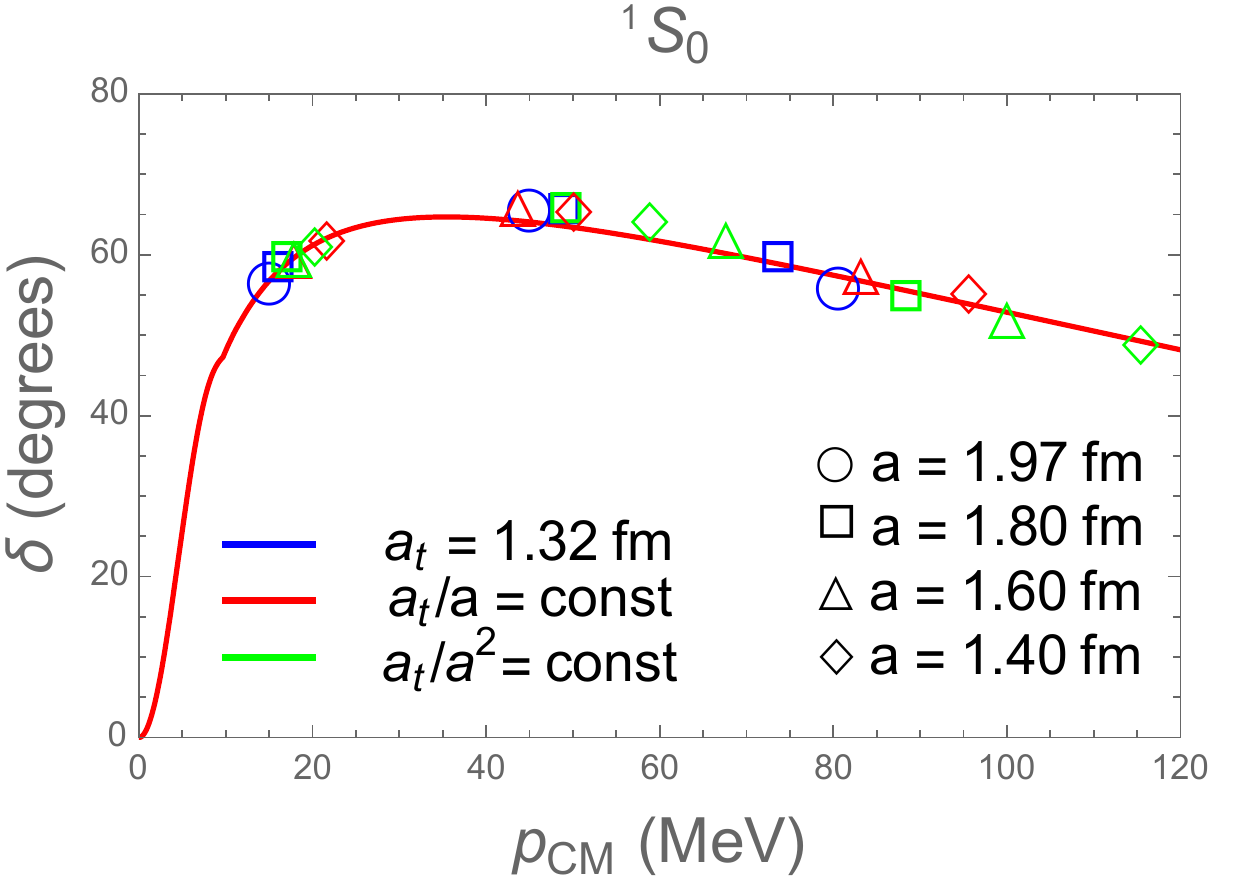}\includegraphics[width=.5\textwidth]{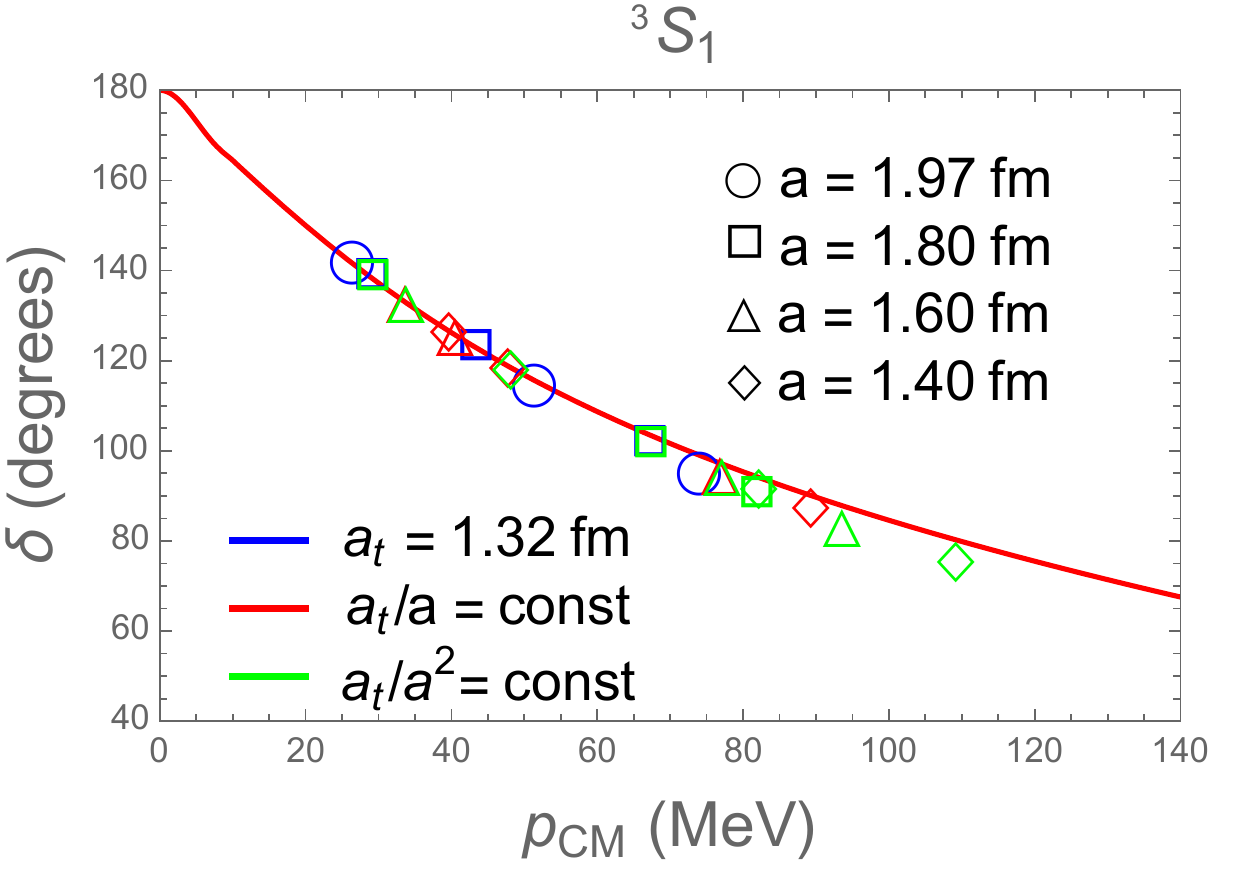}
\caption{Results obtained for the different combinations of $a$ and $a_t$. The shape of the point indicates the value of the spacial spacing, while the color indicates the value of $a_t$ as a function of the value of $a$. In the labeling, "const" means the result of the ratio using the original values of $a$ and $a_t$, i. e., $a = 1.97$~fm and $a_t = 1.32$~fm.  \label{FigSpacingDependence}}
\end{center}
\end{figure}

In Fig.~\ref{FigSpacingDependence} we see how the reduction of the lattice spacings does not change the good description of the $S$-waves. 
Also, a reduction of the spacing allows us reach higher energies, which are still well described by the LO action. 
Regarding the binding energy, we see an equally well (exact) reproduction of the experimental value.
However, the physical value of the LECs and the value of the smearing parameter changes with the spacing. 
This is not surprising for the LECs, since we know that these ones scale, respect to the lattice value, as $C^{phys} = C^{latt}a^2 $.

\begin{table}[h!!]
\centering
\begin{tabular}{|c|c|c|c|c|c|c|}
\hline
      $a$~(fm)  &  $a_t$~(fm)          &  $C_{1S0}$  & $C_{3S1}$ & $b_4$    & $E_B$       \\
       & & (in $10^{-5}$ MeV$^{-2}$) & (in $10^{-5}$ MeV$^{-2}$)& &(MeV) \\
      
\hline
   $1.97$                &   $1.32$                            &   $-4.018$                  &  $-5.865$                & $0.0772$  & $-2.224575$ \\
    $\textcolor{blue}{1.80}$                &   $\textcolor{blue}{1.32}$                                &   $\textcolor{blue}{-3.206}$                  &  $\textcolor{blue}{-4.673}$                & $\textcolor{blue}{0.1064}$ & $\textcolor{blue}{-2.224575}$ \\
    $\textcolor{green}{1.80}$                &   $\textcolor{green}{1.09}$                     &   $\textcolor{green}{-3.357}$                  &  $\textcolor{green}{-4.868}$                & $\textcolor{green}{0.0986}$  & $\textcolor{green}{-2.224575}$\\
   $\textcolor{red}{1.60}$                &   $\textcolor{red}{1.06}$                           &   $\textcolor{red}{-2.532}$                  &  $\textcolor{red}{-3.625}$                & $\textcolor{red}{0.1413}$ &  $\textcolor{red}{-2.224575}$\\
   $\textcolor{green}{1.60}$                &   $\textcolor{green}{0.86}$                   &   $\textcolor{green}{-2.648}$                  &  $\textcolor{green}{-3.776}$                & $\textcolor{green}{0.1320}$  & $\textcolor{green}{-2.224575}$ \\
  $\textcolor{red}{1.40}$                &   $\textcolor{red}{0.93}$                         &   $\textcolor{red}{-1.873}$                  &  $\textcolor{red}{-2.600}$                & $\textcolor{red}{0.2005}$ & $\textcolor{red}{-2.224575}$ \\
   $\textcolor{green}{1.40}$                &   $\textcolor{green}{0.66}$                  &   $\textcolor{green}{-1.975}$                  &  $\textcolor{green}{-2.737}$                & $\textcolor{green}{0.1863}$ & $\textcolor{green}{-2.224575}$ \\
 \hline
\end{tabular}
\caption{\small Best values of the LO LECs and smearing parameter ($b_4$) obtained in the fits with different spacings. The first line corresponds to the standard values of $a$ and $a_t$. The color refers to the value of $a_t$ taken respect to $a$.}
\end{table}

From this study we conclude that, for this moderate reduction of the spacing, the good description of the $np$ $S$-wave phase shift remain the same. 
We will continue further our analysis to reduce the spacing down to $\sim 1$~fm.

\section{Summary and Conclusions}

In this contribution, we showed some recent developments regarding the study of neutron-proton scattering with NLEFT. 
These can be attributed to important improvements in the strategy used to fix the free parameters of the theory. 
The new method allows us to perform the fits in a much more efficient way, reducing considerably the computational time.
This improvement opens the possibility to perform systematic studies of the uncertainties in the determination of the free parameters (not done before due to computational limitations), as well as to study the discretization error of our simulations. 
The latter is crucial in order to explore higher energies.
This will be the goal in future studies, together with the inclusion of higher partial waves. 
To accomplish this task, we are extending our calculation up to N$^3$LO, improving also the LO action to reduce the sign problem in light nuclei simulations.
All these developments will be very important for many-body nuclear simulations in NLEFT, as we plan to show in the near future.

\acknowledgments

I want to thank my collaborators Dechuan Du, Nico Klein, Timo L\"ahde, Dean Lee, Ning Li and Ulf-G. Mei\ss ner for the very valuable discussions during the progress of this work.

\end{document}